\documentclass[a4paper]{article}

\usepackage{eucal}
\usepackage{listings}
\usepackage{xcolor}
\usepackage{multirow}
\usepackage{jheppub}  
\usepackage{dcolumn}
\usepackage[english]{babel}
\usepackage[utf8x]{inputenc}
\usepackage[T1]{fontenc}
\usepackage{palatino}
\usepackage{amsmath}
\usepackage{afterpage}
\usepackage{graphicx, subcaption}
\usepackage[colorinlistoftodos]{todonotes}
\usepackage[colorlinks=true]{hyperref}
\usepackage{makeidx}

\makeindex

\newcommand{\as}{\alpha_{\mathrm{s}}}      

\newcommand{\dif}{\mathrm{d}}

\newcommand{\GeV}{\text{ GeV}}

\newcommand{\kt}{{\boldsymbol k}}
\newcommand{\Lqcd}{\Lambda_{\mathrm QCD}}
\newcommand{\lt}{{\boldsymbol l}}
\newcommand{\MeV}{\text{ MeV}}

\newcommand{\ord}[1]{\CMcal{O}\left(#1\right)}

\newcommand{\plb}[3]{{\it Phys.~Lett.~}{\bf B #1} (#2) #3}

\newcommand{\qt}{{\boldsymbol q}}

\begin{document}

\vspace*{1.2cm}
\thispagestyle{empty}

\begin{center}
{\LARGE \bf Is BFKL factorization valid \\[2mm]
    \hspace{4em} for Mueller-Tang jets?}

\par\vspace*{7mm}\par

{\bigskip \large \bf Dimitri Colferai}
\bigskip
{\large \bf  E-Mail: colferai@fi.infn.it}

\bigskip

{Department of Physics, University of Florence and INFN Florence}

\bigskip

{\it Presented at the Low-$x$ Workshop, Elba Island, Italy, September 27--October 1 2021}

\vspace*{15mm}
\end{center}
\vspace*{1mm}

\begin{abstract}
  TThe perturbative QCD description of high-energy hadroproduction of two hard
  jets separated by a large rapidity gap void of emission (also called
  Mueller-Tang jets) is based on a factorization formula of BFKL type that
  represents exchanges of colour-singlet objects among the external
  particles. This formula resums to all perturbative orders a certain class of
  Feynman diagrams that are supposed to dominate the cross-section in the Regge
  limit. However, the explicit calculations at next-to-leading logarithmic order
  questions the validity of such factorization when an IR safe jet algorithm is
  used to reconstruct jets. We show the origin of such violation of
  factorization, and quantify its impact for LHC phenomenology.  In this
  connection, we estimate the impact of other contributions to the cross-section
  that are not included in the Mueller-Tang factorization formula --- colour
  non-singlet exchanges --- that, at low rapidity separation, compete with the
  singlet ones.
\end{abstract}
 
\section{Introduction}

\begin{figure}[ht]
\begin{center}
  \includegraphics[width=0.35\linewidth]{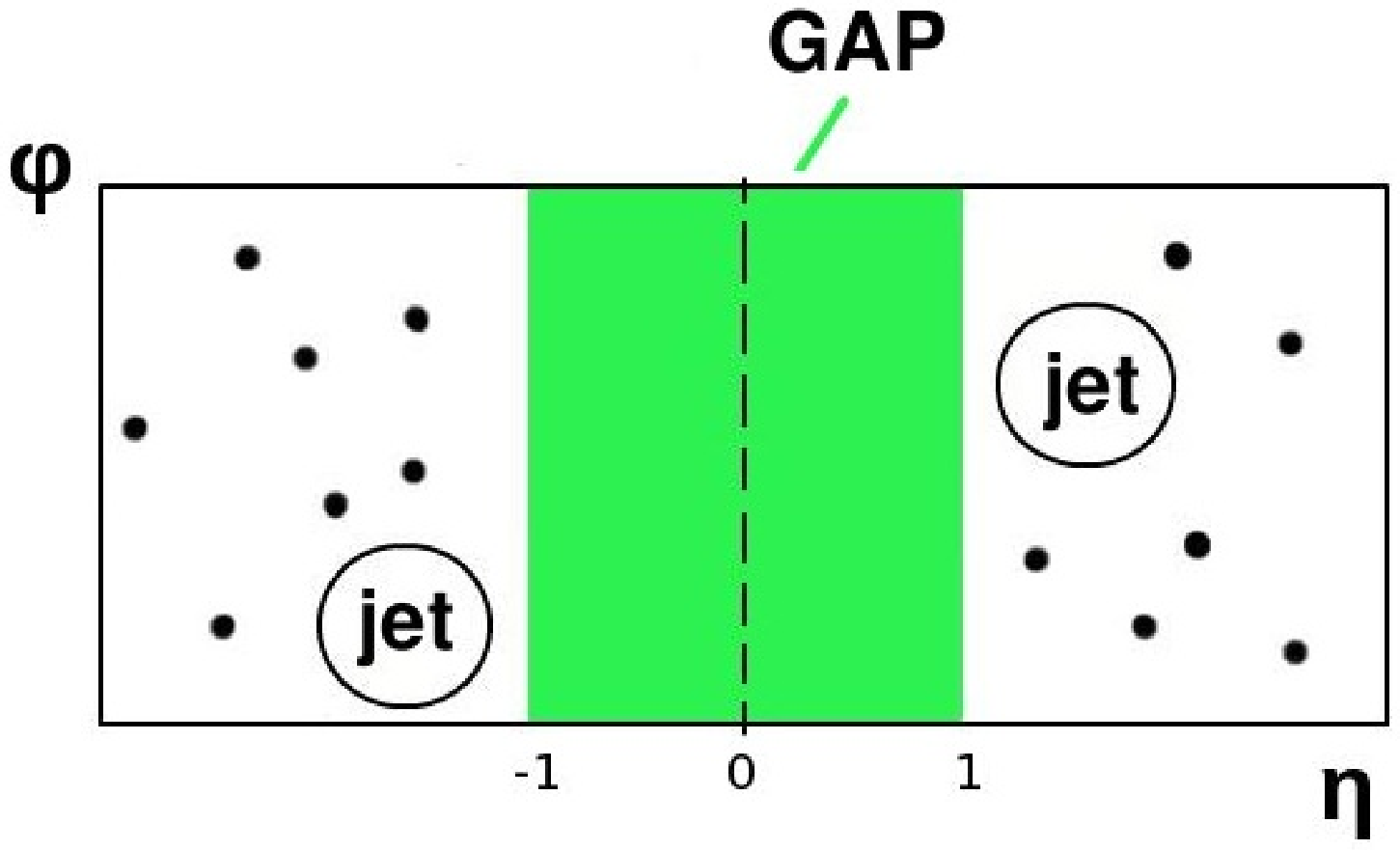}
  \hspace{0.1\linewidth}
  \includegraphics[width=0.35\linewidth]{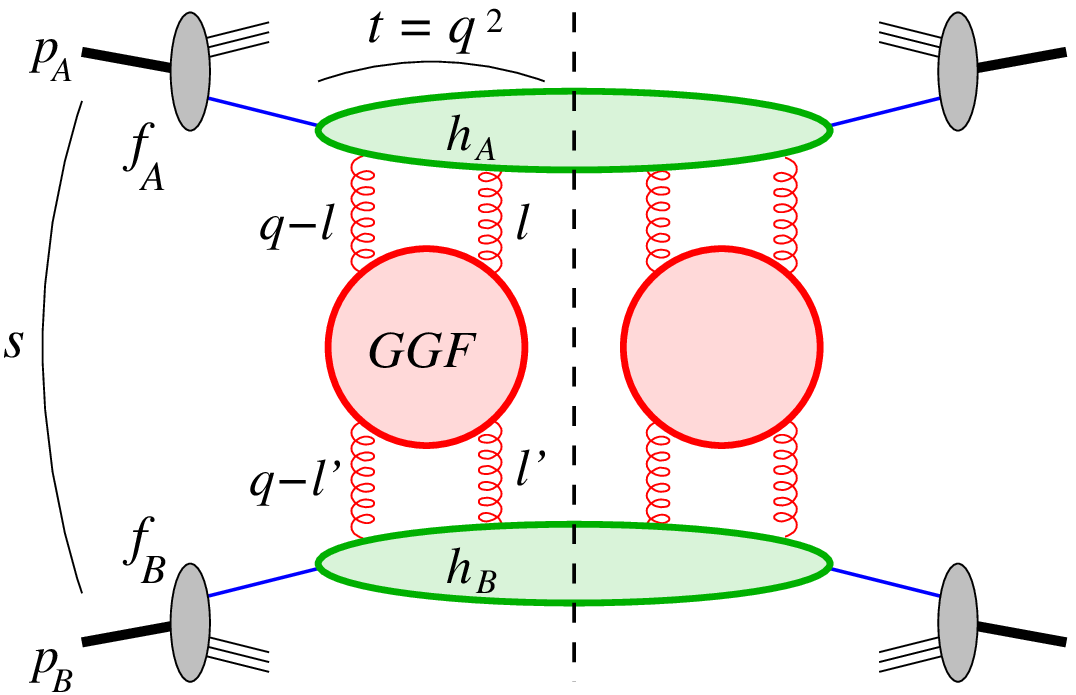}\\
  (a)\hspace{0.45\linewidth}(b)
  \caption{(a) Sketch of particle detection of a jet-gap-jet event in the
    azimuth-rapidity plane. (b) Diagramatic representation or the factorization
    formula for MT jets.}
  \label{f:jgj}
\end{center}
\end{figure}

Mueller-Tang (MT) jets~\cite{MuTa92} are important for studying perturbative
high-energy QCD and the Pomeron at hadron colliders. They are characterized by
final states with at least 2 jets with comparable hard transverse momenta
($\kt_{J1}\sim\kt_{J2} \gg \Lqcd$), well separated in rapidity
$Y\equiv y_{J1}-y_{J2}$, and absence of emission in a given interval of
pseudo-rapidity $\Delta\eta\lesssim Y$ in the central region (the so-called
gap). For this reason, they are also called ``jet-gap-jet'' events, and a
typical final state is depicted in fig.~\ref{f:jgj}a.

The presence of the gap suggests that these events mainly occur when the momentum
exchange between the forward and backward systems is due to a colour-singlet
virtual state: a non-singlet exchange would be characterized most of the times
by final state radiation deposited in the central region.

A large rapidity interval is possible because at LHC the
center-of-mass (CM) energy is much larger than the jet transverse energy.  In
this case, the coefficients of the perturbative series are enhanced by powers of
$Y\simeq \log(s/\kt_J^2)$, and an all-order resummation of the leading terms
$\sim(\as Y)^n$ is needed for a proper determination of the amplitude.

\subsection{Cross section in leading logarithmic approximation}

At lowest perturbative order, a colour-singlet exchange in the $t$-channel is
due to two gluons in colour-singlet combination.  At higher orders, as just
mentioned, the partonic elastic amplitude is affected by powers of
$Y\simeq \log(s/\kt_J^2)$ due to gluon ladder-like diagrams. Such contributions
can be resummed into the so-called BFKL gluon Green function (GGF)~\cite{BFKL}.
It is interesting to observe that such LL loop diagrams are both UV and IR
finite.

By squaring the partonic amplitude, the LL partonic cross-section is then given
by the product of 2 GGFs, which embody the energy-dependence, and two impact
factors (IFs), that couple the gluons to the external particles. In the LL
approximation the IFs are just a trivial product of coupling constants and
colour factors.

Finally, the cross section for MT jets in the LL approximation can be expressed by the
factorization formula (see fig.~\ref{f:jgj}b)
\begin{align}
  \frac{\dif\sigma^{(LL)}}{\dif J_1 \dif J_2} \simeq \int
  \dif (x_1, x_2, \lt_1, \lt'_1, \lt_2, \lt'_2)\; &
  f_A(x_1) \Phi_A(x_1,\lt_1,\lt_2;J_1) G(x_1 x_2 s,\lt_1,\lt_2) \nonumber\\
  &\times G(x_1 x_2 s,\lt'_1,\lt'_2)
  \Phi_B(x_2,\lt'_1,\lt'_2;J_2) f_B(x_2) \;. \label{ff}
\end{align}
Here $J=(y_J,\kt_J)$ represents the set of jet variables,
the GGFs $G$ describe universal gluon dynamics, the IFs $\Phi_i$
describe the coupling of the reggeized gluons or pomerons to the external
particles, and the PDFs $f_i$ describe the partonic content of hadrons.

\subsection{First phenomenological analyses of jet-gap-jet events}

\begin{figure}[ht]
\begin{center}
\includegraphics[width=0.5\linewidth]{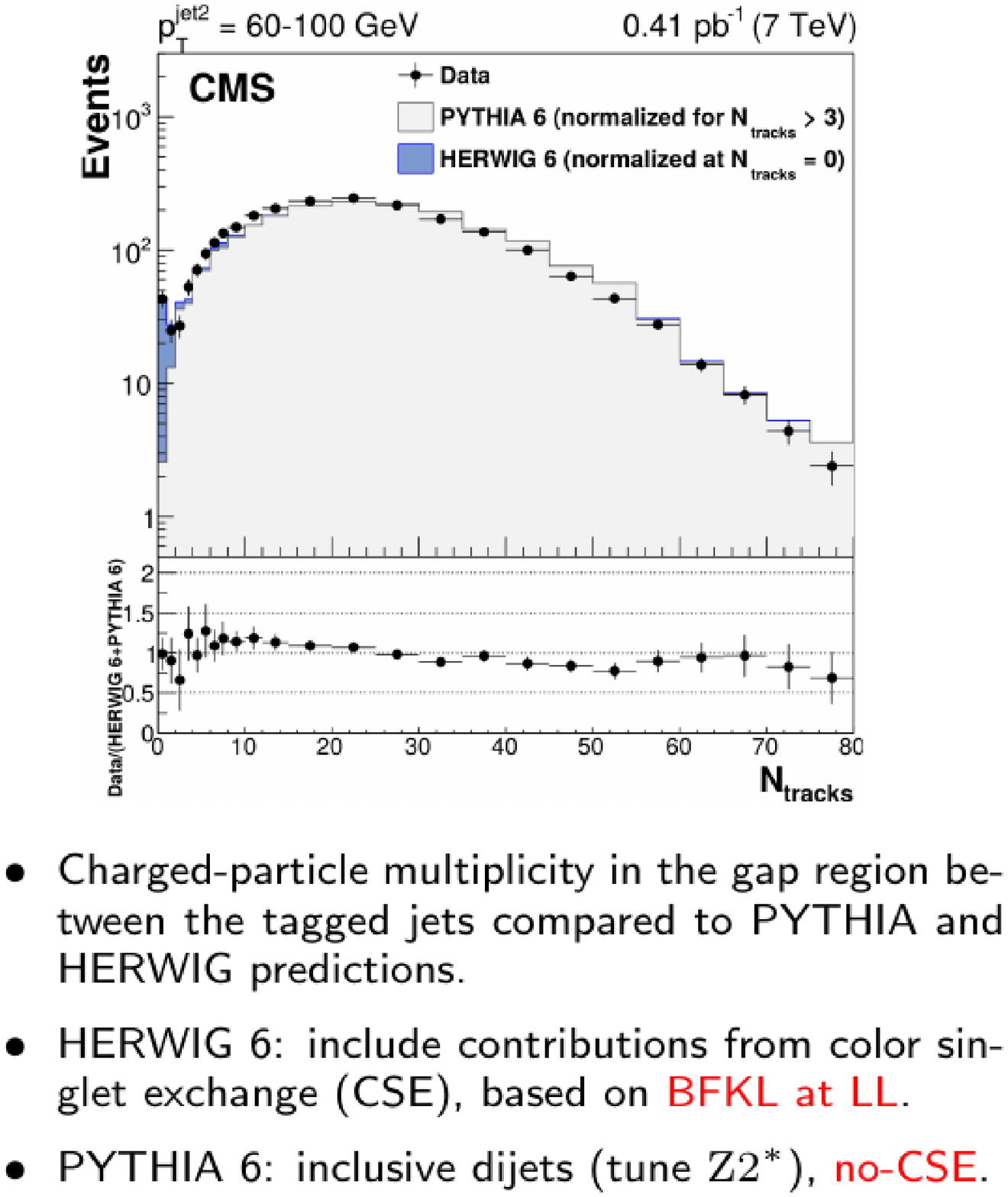}
\caption{CMS measurements of multiplicity of charged particles in the gap
  region, and comparison with Pythia and Herwig predictions.}
\label{f:multiplicity}
\end{center}
\end{figure}

The importance of considering such BFKL contributions to the cross section has
been emphasized since the first analysis by CMS. The plot in
fig.~\ref{f:multiplicity} shows the number of events as a function of the
multiplicity of charged particles in the gap region. We see that both Herwig and
Pythia are able to describe the data if one or more particles are observed
between the jets, but only Herwig agrees in the first bin without observed
particles, and this happens because Herwig includes the contribution of
colour-singlet exchange from BFKL at LL

\begin{figure}[hb]
\begin{center}
\includegraphics[width=0.9\linewidth]{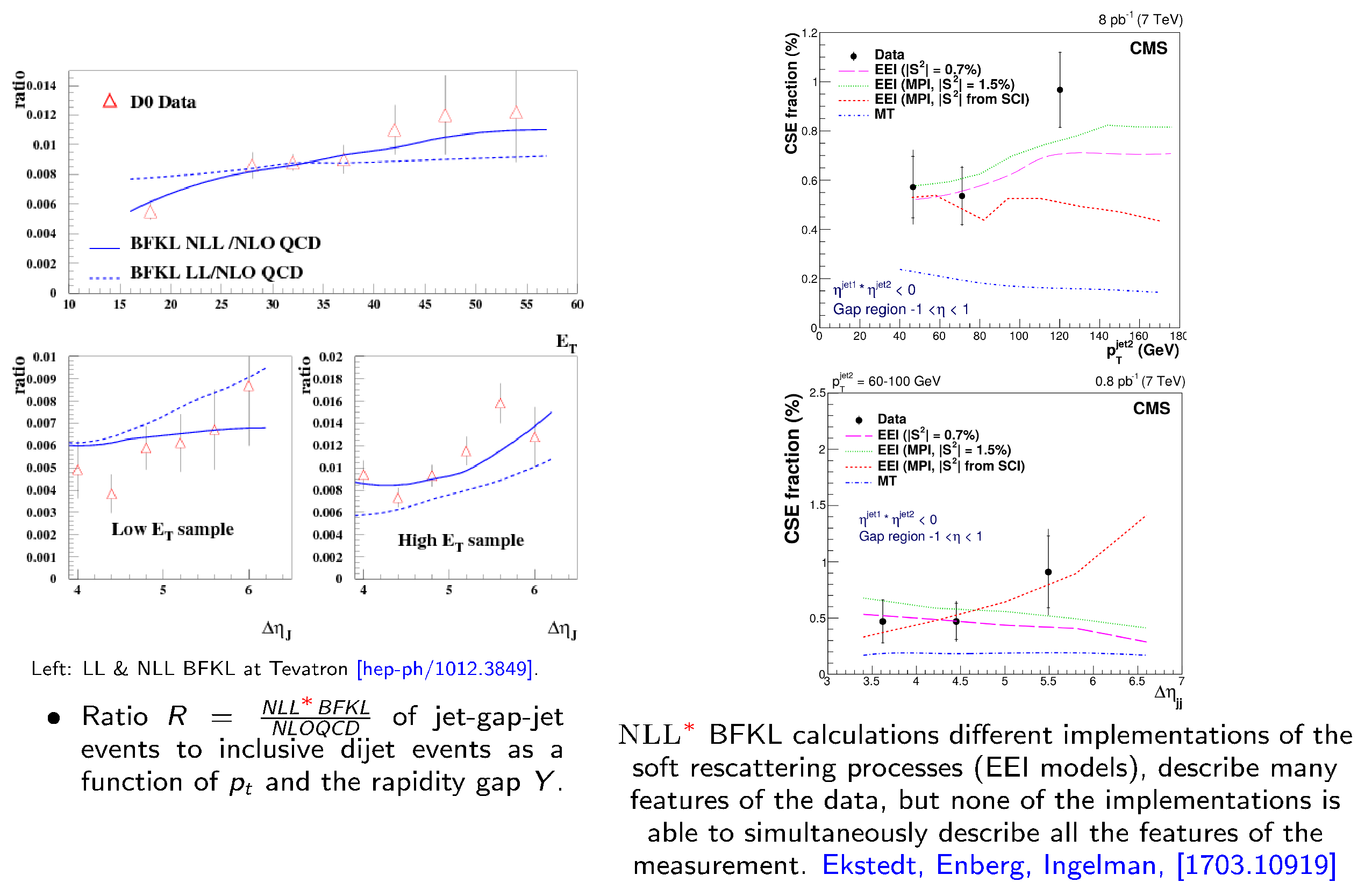}\\
\caption{Comparison of various differential measurements by D0 for jet-gap-jet
  events, and comparison with various theoretical models implementing the BFKL
  GGF in NLL approximation.}
\label{f:d0}
\end{center}
\end{figure}

If one looks at differential distributions of JGJ events, like distributions in
$p_\perp$ or in rapidity distance $Y$, however, the situation is not so
nice. Here LL predictions are unable to describe data (see
fig.~\ref{f:d0}a). Even if one improves the BFKL GGF by adding next-to-leading
logarithmic (NLL) contributions~\cite{NLLFL,NLLCC,KMR10,EEI17}, none of the
implementations is able to simultaneously describe all the features of the
measurements (see fig.~\ref{f:d0}b).

\section{Impact factor in next-to-leading logarithmic approximation}

\subsection{Structure of the calculation and final result}

It appears thus compelling to provide a full NLL description of MT jets. The
idea is to generalize the BFKL factorization formula for MT jets to the NLL
approximation.


The NLL BKFL GGF is known in the non-forward case~\cite{NLLnf}, but due to its
complexity, only the forward version~\cite{NLLFL,NLLCC}, has been used in order
to estimate the contribution of NL logarithmic terms to the cross
section~\cite{KMR10}. However, this is not relevant for our study of the impact
factors.

\begin{figure}[ht]
\begin{center}
\includegraphics[angle=-90,width=0.5\linewidth]{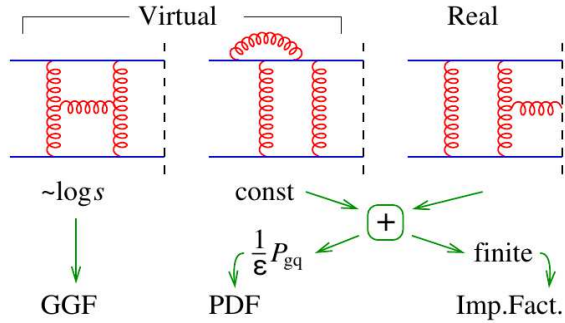}
\caption{Schematics of the decomposition of the (real + virtual) NLO calculation
  for the determination of the NL impact factors.}
\label{f:mtdec}
\end{center}
\end{figure}

The determination of the NL IF can be done with a NLO calculation, which is
affected, of course, by IR (soft and collinear) divergencies.  Actually, the
very existence of NL IF is not a trivial statement. By summing virtual and real
contributions at first perturbative order, one has to prove that
\begin{itemize}
\item the $\log(s)$ terms from virtual corrections reproduce the BFKL kernel
  (and this we already know);
\item the const term of the virtual corrections, which are IR divergent and
  constitutes the virtual part of the IF, when combined with real emission
  terms, must provide a finite remainder, after subtraction of the collinear
  singularities (proportional to the Altarelli-Parisi splitting functions) to be
  absorbed in the PDF;
\end{itemize}
Such finite remainder defines the next-to-leading impact factor. A sketch of
this decomposition is depicted in fig.~\ref{f:mtdec}

\begin{figure}
  \begin{center}
    \includegraphics[width=0.27\linewidth]{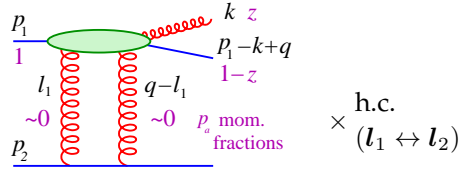}%
    \raisebox{5ex}{
      $\times$
      \begin{minipage}[c]{0.10\linewidth}
      h.c.\\
      $(\lt_1\leftrightarrow\lt_2)$
    \end{minipage}}
    \caption{Kinematics of the calculation of the NL impact factor. Black
      symbols denote 4-vectors, while purple ones denote longitudinal momentum
      fraction.}
    \label{f:kinematics}  
  \end{center}
\end{figure}

The calculation of NL IF for MT jets was performed~\cite{HMMS14a,HMMS14b} using
Lipatov's effective action, and has been confirmed by our independent
calculation~\cite{CoDeRo21}. This is the structure of the result in the case
of incoming quark:
\begin{align*}
  \Phi(\lt_1,\lt_2,\qt)&=\frac{\as^3}{2\pi(N_c^2-1)}\int_0^1\dif z\int\dif\kt\;
  S_J(\kt,\qt,z)\, C_F\frac{1+(1-z)^2}{z}\\
  & \times \left\{C_F^2\frac{z^2\qt^2}{\kt^2(\kt-z\qt)^2}
  +C_F C_A\, f_1(\lt_{1,2},\kt,\qt,z) + C_A^2\, f_2(\lt_{1,2},\kt,\qt)
  \right\}
\end{align*}
It is important to understand the kinematics of the process (see
fig.~\ref{f:kinematics}): after the ``upper'' incoming quark interacts with the
two gluons in colour-singlet, a quark and a gluon can be found in the forward
hemisphere of the final state; the ``lower'' parton $p_2$ remains intact and is
just slightly deflected in the backward hemisphere.

Let me denote with $k$ the outoing gluon momentum, with $\kt$ its transverse
momentum and with $z$ its longitudinal momentum fraction with respect to the
parent quark; $q$ is the overall $t$-channel transferred momentum.  $\kt$ and
$z$ are integration variables. Virtual contributions are contained as
delta-function contributions at $z = 0$ and $\kt = 0$.

We can see the quark-to-gluon splitting function $P_{gq}$ as overall factor, and
then three terms with different colour structures. The integration in the phase
space of the gluon and quark final state has to be restricted by an IR-safe jet
algorithm $S_J$, such as the $k_\perp$-algorithm.

\subsection{Violation of BFKL factorization}

In the diffractive process we are considering, one quark moves in the backward
direction and is identified with the backward jet. The other two partons, whose
distance in azimuth-rapidity is denoted by
$\Delta\Omega = \sqrt{\Delta\phi^2+\Delta y^2}$, are emitted in the forward
hemisphere, so as to produce at least one jet.  There are 3 possibilities:
\begin{itemize}
\item $\Omega < R$ corresponding to a composite jet;
\item $\Omega > R$ where the {\em gluon is the jet} and the quark is outside the
  jet cone;
\item $\Omega > R$ where the {\em quark is the jet} and the gluon is outside the
  jet cone.
\end{itemize}
In the last configuration there is a problem due to the $\dif z/z$ integration
of the $C_A^2\, f_2$ term. In fact, when the quark is the jet, the gluon can
become soft and its phase-space integration is essentially unconstrained.

The limit $z\to 0$ at fixed $\kt$ corresponds to find the gluon in the central
(and backward) region, where the emission probability of the gluon turns out to
be flat in rapidity, and formally the $z$ (or $y_g$) integration diverges.

If we believe the above transition probability to be reliable at least in the
forward hemisphere ($y_g>0$), the longitudinal integration yields a
$\log(s)$. But a $\log(s)$ in the IF is not acceptable, being against the spirit
of BFKL factorization where all the energy-dependence is embodied in the GGFs.

All this looks strange, because one would have argued that gluon emission in the
central region should be dynamical suppressed, due to the singlet exchange in
the $t$-channel. We will solve this puzzle later on. For the moment, we discuss
a proposal to cope with this fact in practice.

In order to avoid this problem, the authors of~\cite{HMMS14a,HMMS14b} impose an
upper limit $M_{X,\max}$ on the invariant mass of the forward diffractive
system. In that case, the $z$ variable is bounded from below and the $z$-integral
is finite.  However, a crucial question then arises: do we really need to impose
a cut on the diffractive mass?  Actually, {\em can} we impose such a constraint?

If it were possible, then one could avoid $\log(s)$ terms in the IF, though at the
price of introducing logarithms of the diffractive mass: $\log(M^2_{X,\max}/\kt_J^2)$.
However, from the experimental point of view, in order to impose such constraint
one should be able to measure the spectator partons, i.e., the proton remnants
(or the intact proton in case of diffractive events).
Since this is not possible with the present experimental detectors at hadron
colliders, other solutions must be found. 

\begin{figure}[ht]
\begin{center}
  \includegraphics[width=0.6\linewidth]{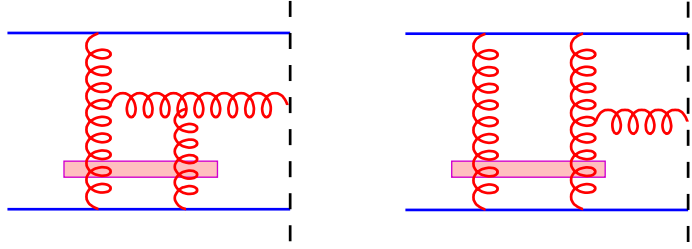}
\caption{Examples of diagrams that involve a non-singlet emission ``above'' the
  emitted gluon, thus producing a $\log(s)$ term in the impact factor. The pink
  rectangles denote colour-singlet projection.}
  \label{f:ec}
\end{center}
\end{figure}

In order to find the origin of such logarithmic contributions in the IF, let us
consider a pair of diagrams contributing to the $C_A^2\,f_2$ term and drawn in
fig.~\ref{f:ec}. It is clear that, if the two $t$-channel (vertical) gluons
emitted by the lower quark are in a colour-singlet state, by colour conservation
the (one or two) upper gluons cannot be in such a state, since a (coloured)
gluon is emitted in the final state. Therefore we cannot claim that this diagram
involves a colour-singlet exchange between the upper and lower system --- the
final state gluon being in the central region.

The option of defining MT jets by selecting those diagrams that involve only
colour-singlet exchanges is not viable, in particular this would end up in a
non-gauge-invariant procedure. We therefore claim that this problem cannot be
avoided and that MT jets are not describable by the naive factorization formula
originally proposed.

At this point, it remains to estimate the size of such violation and possibly to
resum another set of diagrams, if the violation is sizeable.  Given the fact
that we cannot measure particles (partons, hadrons) below some energy threshold
$E_{\mathrm{th}}\sim 200\MeV$ , we can at most require no activity above that
threshold within the rapidity gap.  This prescription is IR safe, because it is
inclusive for gluon energies $E_g < E_{\mathrm{th}}$ (our analysis here proceeds
at partonic level). Since such soft gluons can have arbitrary values of rapidity
between the two jets, we can easily estimate that the logarithmic contribution
to the impact factor is of the order
\begin{equation}
  \Phi_{log} \sim C_A^2 \frac{E_{\mathrm{th}}^2}{\kt_J^2}\log\frac{s}{\kt_J^2}
\end{equation}
Note that this term is regular for $E_{\mathrm{th}}\to 0$, actually it vanishes, at
variance with the $\ord{C_F^2}$ term in the IF which diverges in the same limit.
When evaluated with the values of energies and momenta of typical processes
analysed at LHC, this term turns out to be small, of order $1\%$ or less with
respect to other terms~\cite{CoDeRo21}.

\begin{figure}[ht]
\begin{center}
  \includegraphics[width=0.5\linewidth]{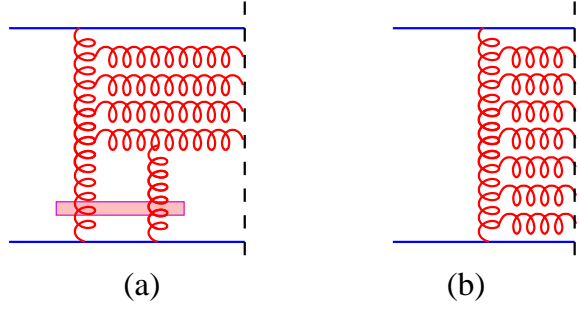}
\caption{Examples of diagrams resumming gluon emission from non-singlet gluon
  exchange. (a) Corrections to the IF; the pink rectangles denote colour-singlet
  projection. (b) Mueller-Navelet-like contributions outside the BFKL
  factorization formula~\eqref{ff}.}
  \label{f:resum}
\end{center}
\end{figure}

Although not really needed from a quantitative point of view at the moment, one
could envisage to resum such logs in the same BFKL spirit, i.e. by considering
diagrams where an arbitrary number of soft (below threshold) gluons are emitted
in the gap (without being detected). This enlarge considerably the number of
diagrams to be taken into account. Some of them, like those in
fig.~\ref{f:resum}a, could be incorporated in the IF, which however acquire a
dependence on the jet rapidity as well as on the gap extension and threshold.

Other diagrams, like those of fig.~\ref{f:resum}b, contribute in a completely
different way, outside the structure of the MT factorization formula. Actually,
they are just diagrams of a Mueller-Navelet (two jet inclusive) process, with
the restriction that the energy of all gluons emitted in the gap region is below
threshold. For threshold energies $E_{\mathrm{th}}\ll |\kt|$, this contribution
can be easily estimated in LL approximation, at least as far as the
$E_{\mathrm{th}}$-dependence is concerned: for soft emissions, virtual and real
corrections cancel each other. Since virtual corrections are always fully taken
into account, while imposing a void gap constrains only real emission, what
remains is essentially equal to the virtual corrections with momenta above
$E_{\mathrm{th}}$. In the LL approximation, virtual corrections are provided by
the exponentiation of the intercept of the reggeized gluon $\omega(\kt)$ with
its internal momentum integrated below $E_{\mathrm{th}}$, resulting in
\begin{align*}
  \omega_{\mathrm{th}}(\kt) &\simeq -\frac{\as N_c}{\pi}\log\frac{|\kt|}{E_{\mathrm{th}}} \\
  \frac{\dif\sigma_{\mathrm{oct}}}{\dif t} &\simeq \frac{\dif\sigma_0}{\dif t}
  \exp\left(-\frac{\as N_c}{\pi}\log\frac{\kt^2}{E_{\mathrm{th}}^2} Y\right)
\end{align*}
to be compared with the MT asymptotic cross section
\begin{align*}
  \frac{\dif\sigma_{\mathrm{sing}}}{\dif t} \simeq \frac{\dif\sigma_0}{\dif t}
  \frac{(\as C_F \pi)^2}{2}\frac{
    \exp\left(\frac{\as N_c}{\pi}8\log2 Y\right)}{[\frac72\as N_C\zeta(3)Y]^3} \;,
\end{align*}
$\dif\sigma_0/\dif t$ being the lowest order (one-gluon exchange) cross section.

\begin{figure}[ht]
\begin{center}
  \includegraphics[width=0.5\linewidth]{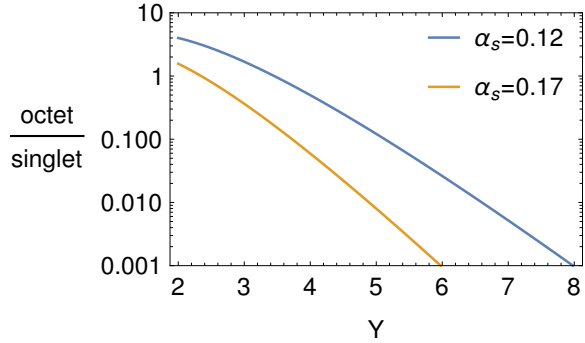}
  \caption{Ratio of differential cross sections in rapidity $Y$ of non-singlet
    (octet) exchanges (fig.~\ref{f:resum}b) emitting gluon having energies below
    threshold and singlet-exchange (fig.~\ref{f:jgj}b). Two values of $\as$ are
    considered.}
  \label{f:singVSoct}
\end{center}
\end{figure}

Such comparison was already made by Mueller and Tang in their original
paper~\cite{MuTa92}, but for very large values of $Y\simeq 12$ and other
parameters not corresponding to LHC kinematics, where the non-singlet exchanges
is strongly suppressed with respect to the singlet ones. Repeating such
comparison with realistic LHC parameters ($\kt= 30 \GeV$, $E_{\mathrm{th}}=0.2\GeV$) 
we find (fig.~\ref{f:singVSoct}) that for $Y\sim 3$ the two contributions are of
the same order, and at $Y\sim 4$ the non-singlet one is still important, about
10\% of the singlet one.

\section{Conclusions}

To summarize, we have demonstrated that, for jet-gap-jet observables, there is
violation of the standard BFKL factorization at NLL level, since the IFs present
logarithmically enhanced energy-dependent contributions. However such terms are
rather small, below 1\% for current measurements of Mueller-Tang jets at LHC,
and their resummation looks not compelling.

Nevertheless, colour non-singlet contributions are expected to be non-negligible
at LHC, in particular for small values of the rapidity distance $Y$ between
jets. Mueller-Navelet contrubution below threshold should in this case be
included, unless NLL corrections to the latter provide a further suppression so
as to render them irrelevant. But this requires further studies.

\vspace{2ex}

Comments: Presented at the Low-$x$ Workshop, Elba Island, Italy, September 27--October 1 2021.

\section*{Acknowledgements}

I thank Krzysztof Kutak and Leszek Motyka for useful discussions on this
subject.\\[1mm]
\noindent
\includegraphics[width=2.2em,angle=90]{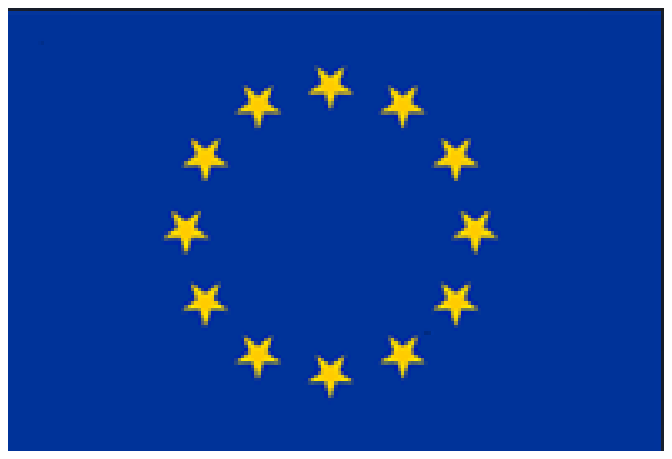}~
\begin{minipage}[b]{0.9\linewidth}
  This project has received funding from the European Union’s Horizon 2020 research and innovation programme under grant agreement No 824093.
\end{minipage}


\begin{thebibliography}{}

\bibitem{MuTa92}
A.~H.~Mueller and W.~K.~Tang,
Phys. Lett. B \textbf{284} (1992), 123-126
doi:10.1016/0370-2693(92)91936-4

\bibitem{BFKL}
  L.~N.~Lipatov,
  Sov. J. Nucl. Phys. \textbf{23} (1976), 338-345;\\
  E.~A.~Kuraev, L.~N.~Lipatov and V.~S.~Fadin,
  [Zh.\ Eksp.\ Teor.\ Fiz.\  {\bf 72} (1977) 377]
  ;\\
  I.~I.~Balitsky and L.~N.~Lipatov,
  Sov.\ J.\ Nucl.\ Phys.\  {\bf 28} (1978) 822.

\bibitem{NLLnf}
V.~S.~Fadin and R.~Fiore,
Phys. Rev. D \textbf{72} (2005), 014018.

\bibitem{NLLFL}
V.S.~Fadin and L.N.~Lipatov, \plb{429}{1998}{127}.\\

\bibitem{NLLCC}
G.~Camici and M.~Ciafaloni, \plb{386}{1996}{341};
\plb{412}{1997}{396}, [Erratum-ibid.\plb{417}{1997}{390}];
\plb{430}{1998}{349}.

\bibitem{KMR10}
O.~Kepka, C.~Marquet and C.~Royon,
Phys. Rev. D \textbf{83} (2011), 034036
[arXiv:1012.3849 [hep-ph]].

\bibitem{EEI17}
A.~Ekstedt, R.~Enberg and G.~Ingelman,
[arXiv:1703.10919 [hep-ph]].

\bibitem{HMMS14a}
M.~Hentschinski, J.~D.~Madrigal Mart\'\i{}nez, B.~Murdaca and A.~Sabio Vera,
Nucl. Phys. B \textbf{887} (2014), 309-337
[arXiv:1406.5625 [hep-ph]].

\bibitem{HMMS14b}
M.~Hentschinski, J.~D.~M.~Mart\'\i{}nez, B.~Murdaca and A.~Sabio Vera,
Nucl. Phys. B \textbf{889} (2014), 549-579
[arXiv:1409.6704 [hep-ph]].

\bibitem{CoDeRo21}
  D.~Colferai, F.~Deganutti and C.~Royon, in preparation.
  
\end{thebibliography}
\end{document}